\documentclass{article}

\usepackage{url}
\usepackage{amsmath}
\usepackage[margin=1in]{geometry}
\usepackage{bm}
\usepackage[affil-it]{authblk}
\usepackage{float}
\usepackage{graphicx}

\renewcommand{\texttrademark}{\textsuperscript{\tiny TM}}

\newcommand{\citep}{\cite}
\newcommand{\citet}{\cite}

\title{Bayesian data assimilation provides rapid decision support for vector-borne diseases}
\date{}
\author{Chris P Jewell}
\affil{CHICAS\\ Lancaster University\\ Bailrigg\\ Lancaster\\ LA1 4YG\\ UK}
\author{Richard G Brown}
\affil{Institute of Fundamental Sciences\\ Massey University\\ Private Bag 11222\\ Palmerston North 4442\\ New Zealand}

\begin{document}

\maketitle

\begin{abstract}
Predicting the spread of vector-borne diseases in response to incursions requires knowledge of both host and vector demographics in advance of an 
outbreak. Whereas host population data is typically available, for novel disease introductions there is a high chance of the pathogen utilising a vector 
for which data is unavailable. This presents a barrier to estimating the parameters of dynamical models representing host-vector-pathogen 
interaction, and hence limits their ability to provide quantitative risk forecasts. The \emph{Theileria orientalis} (Ikeda) outbreak in New Zealand cattle 
demonstrates this problem: even though the vector has received extensive laboratory study, a high degree of uncertainty persists over its national 
demographic distribution. Addressing this, we develop a Bayesian data assimilation approach whereby indirect observations of vector activity inform 
a seasonal spatio-temporal risk surface within a stochastic epidemic model. We provide quantitative predictions for the future spread of the 
epidemic, quantifying uncertainty in the model parameters, case infection times, and the disease status of undetected infections. Importantly, we 
demonstrate how our model learns sequentially as the epidemic unfolds, and provides evidence for changing epidemic dynamics through time. Our 
approach therefore provides a significant advance in rapid decision support for novel vector-borne disease outbreaks.
\end{abstract}


\section{Introduction}

During outbreaks of infectious diseases, effective decision making is key to implementing efficient control measures.  Quantitative methods are now commonplace for supporting such decisions, and in particular mathematical models are now central to informing control strategy \citep{RocEtAl14}.  Before a disease outbreak, models may be used offline to investigate disease dynamics within a particular population, and hence inform policies such as childhood immunisation \citep{GrEtAl01} and livestock disease containment \citep{TilEtAl06}.  During an outbreak, models have the potential to be used for forecasting future disease spread, provided the difficult task of adequate design and calibration is performed appropriately \citep{WearEtAl05}.  In an era characterised by rapid climatic and sociological change, the increasing emergence of new diseases therefore requires a modelling approach that not only adapts quickly to the current outbreak, but is also flexible to the availability of data \cite{JonEtAl08}.

For contagious diseases such as SARS and foot and mouth disease, recent methodology has enabled models to be employed in real-time outbreak situations \citep{CaucEtAl06, JewEtAl09c}.  At any particular time-point during the epidemic, a forecast of ongoing disease spread may be calculated given knowledge of the underlying population and case data collected so far.  Such a forecast may be used not only for monitoring the overall extent of the outbreak, but also for optimal targeting of control strategies, such as surveillance and quarantine, to high risk individuals \citep{JewEtAl2009a}.  However, mathematical models of vector-borne diseases do not appear to have been adopted for real-time forecasting purposes, despite their long history (see for example \citep{Porco1999,ManSarSin11,Sutton2012,Lourenco2014}).  A possible reason for this is that whilst high quality host demographic data are often available from government databases, less is known about national distributions of vector populations.  Indeed, keeping current ecological records on vector populations is subject to prioritisation, such that high quality information relevant to all possible vector-borne diseases is not guaranteed \citep{BrakEtAl11}.  Furthermore, even where such records do exist, the effect of climate change on vector populations is likely to increase the rate at which existing ecological vector studies become
irrelevant to new outbreaks \citep{Purse2005,Bouzid2014,OsBr15}.  Thus a particular challenge for forecasting is to construct a model which is able to adapt quickly to the spatial and seasonal characteristics of a new vector ecology, without having detailed ecological data \citep{Graesboll2014,Parry2014}.

An important aspect of forecasting is the capacity to draw information from all available sources of data. Fitting models to data in this way allows inference about unknown model parameters, quantifies uncertainty about missing data, and aids choice between competing model structures \citep{LuoEtAl11}.   As previously discussed in Jewell et al.\citep{JewEtAl09c}, a Bayesian approach to data assimilation and forecasting offers a number of advantages for prediction over classical approaches.  Firstly, the
probabilistic likelihood-based framework allows a highly flexible approach for assimilating a wide variety of
 available information in a way that allows the model to adapt to the availability of data \citep{PrEtAl10,Jewell2012}.
 Secondly, the ability to use data augmentation Markov-chain Monte Carlo methods to sample from a posterior distribution provides the opportunity to treat missing data, such as unobserved infection times, as latent variables even in large populations \citep{NealRob04}.   Finally, the forecast itself
is represented by the Bayesian predictive distribution -- the
probability distribution of the future epidemic, conditional on what has been
observed to date as well as the epidemic model structure.  Importantly for decision making, the predictive distribution rigorously quantifies uncertainty both in terms of the stochasticity of the epidemic process and parameter estimation \citep{Dawid1984}.

In this paper, we present for the first time a fully Bayesian approach to inference and forecasting for vector-borne diseases in the absence of detailed information on vector ecology.   Motivated by an outbreak of a novel tick-borne pathogen in New Zealand cattle, we construct an epidemic model that represents heterogeneity in both the host and vector populations.  Most significantly, we capture spatio-temporal variation in disease transmission due to vector abundance using a seasonal discrete-space latent risk surface, informed by indirectly collected serosurveillance data, heuristic expert opinion, and the case timeseries itself.  A trans-dimensional Markov-chain Monte Carlo algorithm is used to fit the model to the available data at various timepoints throughout the outbreak, from which we present forecasts of ongoing disease spread.

\section{Motivating example}
This study was motivated by a recent
epidemic of theileriosis in New Zealand cattle, caused by the vector-borne pathogen
\emph{Theileria orientalis} (Ikeda) \citep{LawEtAl2013}. Whilst many \emph{T. orientalis} genotypes
are endemic in NZ, and cause only rare cases of clinical disease, the Ikeda
genotype appears to be associated with high morbidity and
mortality haemolytic anaemia \citep{McFadden2011,Islam2011}. The tick
\emph{Haemaphysalis longicornis} is a putative vector for \emph{T. orientalis}
spp. in New Zealand, and is known to be endemic throughout
the North Island \citep{JamEtAl84}. This vector is sensitive to
climatic conditions, and therefore has a spatially varying distribution as well
as a seasonal pattern of activity \citep{Heath81}. Alternative vectors have been
hypothesised, in particular the \emph{Stomoxys calcitrans} stablefly, which is
active in the same regions as the tick \citep{Heath02}. Beyond laboratory
studies, however, little is known about the quantitative relationship between
climatic factors and vector abundance in the environment. The likely
determinants of disease
spread are therefore environmental factors related to vector presence, distance
from infected herds, and animal movements. 

\subsection{Case report data}
\begin{figure}
\centering
\includegraphics[width=0.6\textwidth]{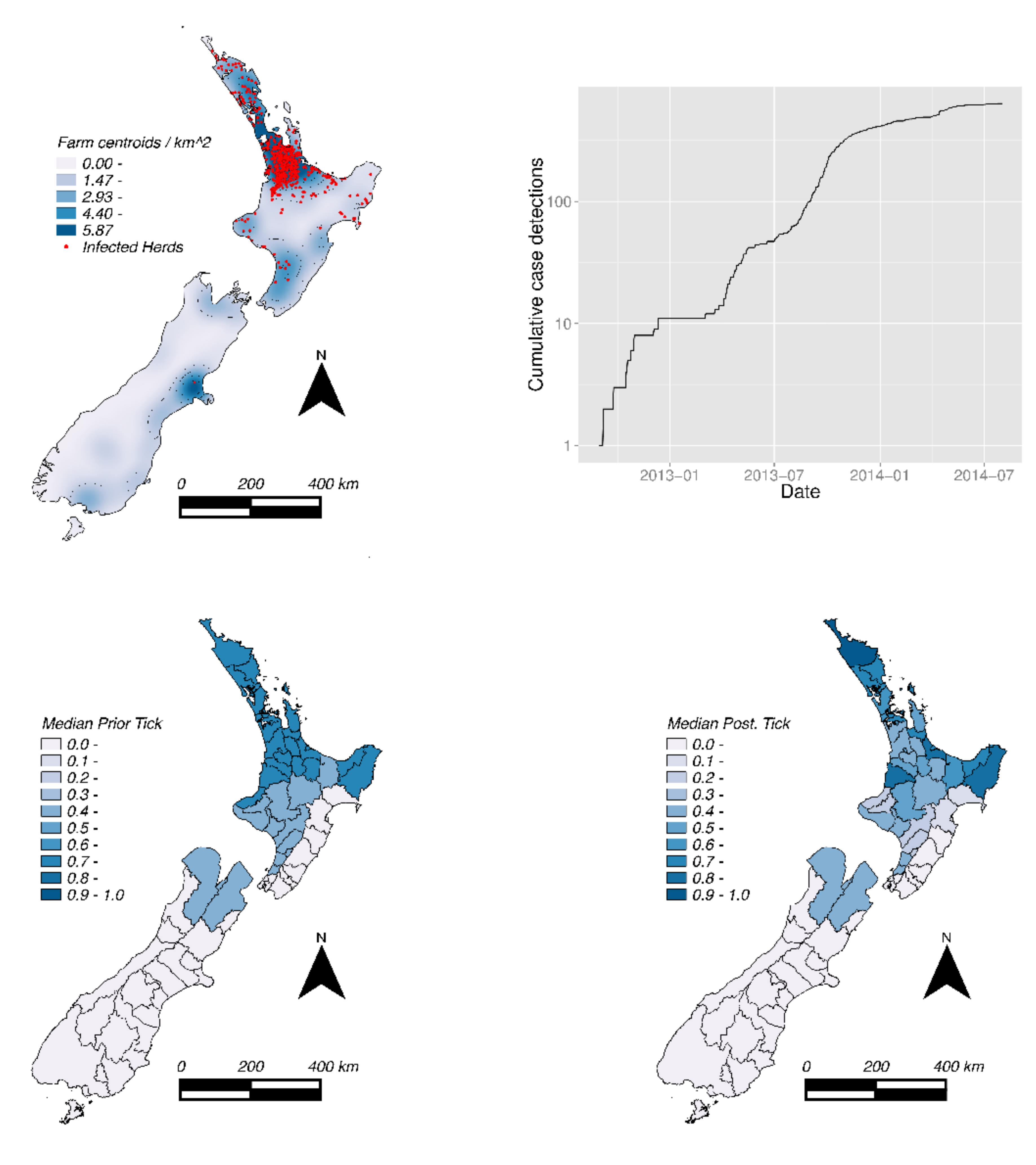}
\caption{\label{fig:Cases-spatial}Summary of the New Zealand \emph{T. orientalis} Ikeda
epidemic as of 1st August 2014.  Top row: Spatial distribution of infected herds, with the log
cumulative number of case detections throughout the epidemic. Bottom row: Median
prior and posterior tick occurrence probabilities for NZ TLAs.}
\end{figure}

The first case of bovine theileriosis in NZ due to \emph{T. orientalis} (Ikeda) was 
detected on the 12th August 2012 \citep{McFad2013,LawEtAl2013}. By 1st August 2014, the outbreak had
grown
to 633 infected cattle herds, concentrated
largely in the Waikato region of the North Island. Figure \ref{fig:Cases-spatial} shows the
spatial distribution of cases as well as the log cumulative case detections
timeseries. The latter
demonstrates a pronounced seasonality, with periods of low case incidence
corresponding
to winter and summer, and higher incidence in autumn and spring. This is likely
due to the seasonality of the putative vector populations, though increased
physiological stress experienced by cattle post-calving in late winter
(August-September) may also have an effect.

The case data is curated by AsureQuality's Agribase{\texttrademark} database.
Each case is assigned a unique identifier and a detection time. Cases are joined
to the Farms OnLine demographic data (see below) using
spatial queries based on farm geospatial polygon records.  

\subsection{Demographic Data}
Demographic data characterising the NZ cattle population were obtained
from Farms OnLine (FOL), the New Zealand government-owned database 
of rural properties, as well as from the National Animal Identification and
Tracking (NAIT) cattle movement database.

The FOL data comprised a list of 220668 premises each with a unique identifier
(FOL ID), geographic centroid using the New Zealand Transverse Mercator
projection (EPSG:2936), and the number of beef and dairy cattle.
To identify those farms owning cattle, we limited our working dataset to
included herds with IDs present in the NAIT movement records (see below),
or being listed as having at least one beef or dairy animal, or having been
detected
as a \emph{T. orientalis} (Ikeda) case. The resultant working dataset
contained 100288 herds.

Cattle movement data was supplied from NAIT in the form of 611230 animal
movement records spanning the 557 day period from 1st January 2012 to 31st
July 2013.  Each record represents a cohort of cattle moved and includes
the date, source and destination identifier for the respective Persons In
Charge of Animals (PICA), and number of animals moved.  The NAIT data is
represented by a geolocated dynamic network, with nodes corresponding to
the 100288 herds in the FOL dataset, and directed edges weighted by the
frequency of animal movements (see supporting information). In total,
520940 non-zero directed edges were identified, representing 0.005\% of the
maximum possible edges in the network (i.e. $100288^{2}$).

\subsection{Spatio-temporal vector distribution\label{sec:Tick-distribution}}
In the absence of direct observations, data on spatial vector abundance on a 
national scale were available in the form of expert entomological opinion and indirect 
observations of herds exposed to any \emph{T. orientalis} genotype.  These were supplied as areal aggregations for
72 Territorial Land Authority (TLA) regions across NZ.

Expert opinion on the spatial distribution of the putative vector
\emph{Haemaphysallis longicornis}
was obtained from Dr Allen Heath, encoded by classifying TLAs into high, medium, and
low tick risk (Figure \ref{fig:Cases-spatial}).  Laboratory studies of Ixodid tick species indicate threshold effects of both humidity and temperature on 
development and activity.  For example,  \citet{St94} demonstrate greatly increased mortality below approximately 93\% relative humidity, and 
\citet{OgEtAl04} conclude a sharp-shouldered power law curve for reproductive activity in response to temperature.  These findings are consistent 
with both \citet{Heath81} and \citet{KnRud82}, suggesting steeply increasing and decreasing tick activity in response to seasonal climatic variation.

After the discovery of the \emph{T. orientalis} Ikeda outbreak, stored blood samples that
had been collected from NZ cattle herds in conjunction with routine BVD
surveillance
were PCR tested for the presence of \emph{T. orientalis} subspecies \citep{PerEtAl2015, PulEtAl2015}.  
These data provide the number of farms tested and the number 
of farms returning positive for \emph{T. orientalis} in each TLA region.
 These data allow us to calculate the apparent prevalence
of endemic \emph{T. orientalis} strains
which serves as a indirect measure of vector activity, without having to
specifically identify the vector species.

\section{Modelling}

\subsection{Epidemic model}
The occurrence of cases of \emph{T. orientalis }(Ikeda) in the NZ
cattle population is represented by a continuous time SID model, 
which assumes that herds progress from Susceptible
to Infected, and are subsequently Detected according to a 
time inhomogeneous Poisson process \citep{JewEtAl09c}. In contrast to other
epidemic
scenarios, no ``removed'' status is used, since we assume that once
\emph{T. orientalis} Ikeda is circulating within the herd, the herd remains
infectious. In this sense, our basic model setup resembles
an SI model, with infection times being indirectly observed via
detection events. 

The effects of spatial location, cattle breed, NAIT-recorded animal movements,
tick
density and seasonality are captured by specifying a model for the
pairwise disease transmission rate. We assume that at time $t$, a susceptible
herd $j\in\mathcal{S}(t)$ experiences infectious pressure
at rate
\begin{equation}\label{eq:lambdaj}
\lambda_j(t) = \sum_{i\in\mathcal{I}(t)} \beta_{ij}(t) 
\end{equation}
where $\mathcal{S}(t)$ and $\mathcal{I}(t)$ are the sets of susceptible and infected herds at time $t$ respectively. Further, we assume
that infection
is transmitted between an infected herd $i$ and susceptible $j$ at time
$t$ at rate
\begin{equation}\label{eq:betaij}
\beta_{ij}(t) =
h(j,t;\bm{\psi})\left[\beta_{1}K(i,j;\delta)+\beta_{2}c_{ij}\right]
\end{equation}
where $h(j,t;\bm{\psi})\geq 0$ represents susceptibility of farm $j$ at time
$t$. $K(i,j;\delta)$ is a
function describing the decay of transmissibility with distance between
herds for non-network infections, with $\beta_{1}$ the baseline rate
of disease transmission. The daily frequency of animal movements between
farms $i$ and $j$ is denoted $c_{ij}$, with parameter $\beta_{2}$
interpreted as the maximal probability of an animal movement resulting
in infection.

We define 
\[
K(i,j;\delta)=\frac{\delta}{\left(\delta^{2}+||x_{i}-x_{j}||^{2}\right)^{\omega}}.
\]
 a Cauchy-type decay kernel with distance
$||x_{i}-x_{j}||$ between locations $x_{i}$ and $x_{j}$, and decay parameter $\delta$.  
$\omega=1.2$ is chosen to optimise statistical identifiability
between $\beta_{1}$ and $\delta$.

Given the paucity of quantitative data relating vector activity to
measurable spatial and temporal climatic data, as well as uncertainty in
breed-susceptibility to \emph{T. orientalis} Ikeda, we model seasonal infection
risk as a separable spatiotemporal latent process
\begin{equation}
h(j,t;\bm{\psi})=s(t;\bm{\alpha},\nu)\zeta^{\kappa_j} p_{k(j)}
\label{eq:hFunction}
\end{equation}
where $\bm{\psi} = \{\bm{\alpha}, \nu, \zeta, \bm{p}\}$.  $s(t;\bm{\alpha},\nu)$ represents the seasonal transmission
risk at time $t$, and $\zeta$ represents the susceptibility of dairy farms relative
to non-dairy; $\kappa_j$ is 1 if $j$ is a dairy farm and 0 otherwise.
The probability parameter $p_{k(j)}$ is a proxy for vector occurrence on
farm $j$, a member of TLA region $k$, and allows us to connect the
epidemic model to independent \emph{T. orientalis} surveillance.

The biannual pattern of theileriosis incidence indicates the need for a flexible seasonal function
capable of capturing differences in transmission peaks and troughs throughout the year. The observed threshold effects of humidity and temperature 
on tick activity suggests that a steep-shouldered function approximated by a square wave might be appropriate.  However, the combined effect of 
humidity and temperature on the vector, the degree of uncertainty surrounding the identity of other vector species, and the fact that this model is 
capturing the occurrence of new cases rather than the vector itself, might suggest a sinusoidal function to be more appropriate \citep{Parry2014}.  To 
address this, we first adopted a piecewise cubic spline function as an analytically tractable approximation to a trigonometric function (see supporting 
information).  Surprisingly, on the basis of in-sample predictive performance (Figure \ref{fig:ispred}) this function was rejected in favour of a periodic 
square wave function
with changepoints fixed at quarter-year epochs:
\begin{equation}\label{eq:seasonal}
s(t; \bm{\alpha}, \nu) = 
\begin{cases}
  1 & \mathrm{if }\; 0 \leq t^\star < 0.25 \\ 
  \alpha_1 & \mathrm{if }\; 0.25 \leq t^\star < 0.5 \\
  \alpha_2 & \mathrm{if }\; 0.5 \leq t^\star < 0.75 \\
  \alpha_3 & \mathrm{if }\; 0.75 \leq t^\star < 1
\end{cases},
t^\star=t+\nu - \left\lfloor t+\nu \right\rfloor
\end{equation}
with $t$ in years. Setting the height of the autumn peak to 1 enables
identifiability between the $\bm{\alpha}$ and
$\bm{\beta}=\{\beta_1,\beta_2\}$, and we allow $0 \le \nu \le 0.5$ to
allow fine tuning of the phase of the seasonal function.
 
We define the infectious period $d$ to be the time between an infection and
detection.
We assume that for each individual $i$, the infectious period is conditionally
independent
given the infection times, and distributed according to 
\[
d_i\sim \mbox{Gamma}(a,b)
\]
with $a=4$ based on previous infectious disease analyses (see for example
\citet{JewEtAl09c}), and $b$ an unknown scale parameter.

\subsection{Surveillance model}
The specification of $p_k$ as the occurrence probability for ticks in TLA region
$k$ above presents the opportunity to model an independent disease testing
process alongside the epidemic. From samples obtained from BVD surveillance, we
have for each TLA region $k$ the number of herds tested, $n_k$, and number
testing positive for \emph{T. orientalis} species, $x_k$. We assume a Binomial
model such that
\begin{equation}
x_k \sim \mbox{Binomial}(n_k, p_k) \label{eq:survModel}
\end{equation}
allowing us to make inference on $p_k$ for each TLA region. We remark that $p_k$
is only a proxy for vector occurrence, since the link between number of cases
testing positive and vector activity is complicated by many factors including
the exposure of the host to the vector, host genetics, and test sensitivity.
Since \emph{T. orientalis} requires a vector for transmission between cattle hosts, $p_k$ may
then be thought of as a measure of the risk that infection will spread through a
herd, given that it is introduced.

\section{Data assimilation and model fitting}

In this section we describe in outline how the epidemic and surveillance models
may be used together to estimate the joint posterior distribution of the model
parameters, infection times of detected herds, and the presence of undetected
infections.  This then facilitates the calculation of the predictive distribution of the epidemic.
The implementation for model fitting and simulation is available as an
R package at \url{https://github.com/chrism0dwk/infer/releases/tag/nztheileria-v1.0}.

We proceed by assuming the epidemic process and BVD surveillance programme to be
independent. This allows us to multiply the statistical likelihood functions for
the epidemic and surveillance models -- $L_E(\bm{\theta}, \bm{I} | \bm{D})$ and
$L_S(\bm{p} | \bm{X})$ respectively -- to obtain a joint likelihood function for
parameters $\bm{\theta} =\{\bm{\alpha}, \beta_1, \beta_2, \nu, \zeta, \bm{p},b\}$ and
unobserved (and undetected) infection times $\bm{I}$, given the case detections
$\bm{D}$ and surveillance data $\bm{X}$ (see supporting information for full details).

The joint posterior distribution function $\pi(\bm{\theta}, \bm{I} | \bm{D},
\bm{X})$ is proportional to the product of the joint likelihood function and
prior distributions $f_\theta(\theta)$ for each parameter
\begin{equation}
\pi(\bm{\theta}, \bm{I} | \bm{D}, \bm{X}) \propto L_E(\bm{\theta}, \bm{I} |
\bm{D})L_S(\bm{p} | \bm{X})\prod_{\bm{\theta}} f_\theta(\theta)
\end{equation}
which allows inference on tick occurrence probabilities $\bm{p}$ to be informed
by both the epidemic data \emph{and} the BVD sampling data. $L_E(\bm{\theta} |
\bm{I}, \bm{D})$ takes the form of a continuous time inhomogeneous Poisson
process likelihood where individuals become infected according to an exponential
distribution with rate given by the infectious pressure $\lambda_j(t)$ from
Equation \ref{eq:lambdaj}.
$L_S(\bm{p} | \bm{X})$ then takes a Binomial likelihood function for 
independent observations from each TLA region.

Prior probability distributions are chosen for each parameter, informed by
expert opinion and heuristic expectation of the resultant epidemic (see
supporting information). The latter was obtained by simulation exploration
of the behaviour of the model without consideration of the case detection
timeseries to date.  To incorporate expert opinion on spatial tick
distribution, independent Beta($a_{k},b_{k}$) prior distributions are
chosen to reflect ``high'', ``medium'', and ``low'' risk, and are applied
to $p_{k}$ for each region corresponding to the expert-classified TLA
regions. The properties of the Beta distributions chosen are shown in the
supporting information.

The Bayesian model was fitted to the observed case data using a
modification of the adaptive reversible jump Markov chain Monte Carlo
algorithm presented in \citet{JewEtAl09c}. This algorithm performs
inference by drawing samples from the joint posterior distribution over the
model parameters, infection times, and occult infections. In our particular
implementation, we use ASIS methodology to enable the algorithm to work
efficiently in the face of strong \emph{a priori} dependence between the
marginal posterior distributions for the infection times and infectious
period scale parameter $b$ \citep{Yu2011}.  Convergence of the algorithm was
confirmed by running 4 parallel independent chains starting at randomly chosen values, as shown in the supporting information. 
 
Having estimated the joint posterior distribution, we employ a continuous time Doob-Gillespie simulation algorithm to construct the posterior 
predictive distribution of the future epidemic, with retrospective sampling
used to account for the seasonal function \citep{JewEtAl09c}. Whilst being less
computationally efficient than a discrete time algorithm, this approach
avoids discretisation error which might bias the resulting disease forecast. The
predictive distribution $f_Y(Y | \bm{D}, \bm{X})$ of the future epidemic $Y$
conditional on the observed case detections and surveillance data is calculated
by simulating over the joint posterior distribution such that
\begin{equation}\label{eq:predDistn}
f_Y(Y | \bm{I}, \bm{D}, \bm{X}) = \int_{\bm{\Theta},\bm{I}} f_Y(Y | \bm{D},
\bm{X}, \bm{\theta}, \bm{I}) \pi(\bm{\theta}, \bm{I} | \bm{D}, \bm{X})
d\bm{\theta}d\bm{I}
\end{equation}

\section{Results}
The analysis of the New Zealand \emph{Theileria orientalis} (Ikeda) outbreak
began
on the 1st November 2013, once it became clear that the epidemic had
established.
Ongoing predictive analyses were made on a monthly basis as updates to the case detection dataset were obtained. We summarise these results at 
quarterly
intervals --
1st November 2013, 1st February 2014, 1st May 2014, and 1st August 2014 --
indicating
how the predictions adapt in the face of learning from increasing case data.
Here we focus on
results relevant to forecasting, though further results relevant to now-casting may be found in the supporting information.

\begin{figure}
\centering
\includegraphics[width=0.9\textwidth]{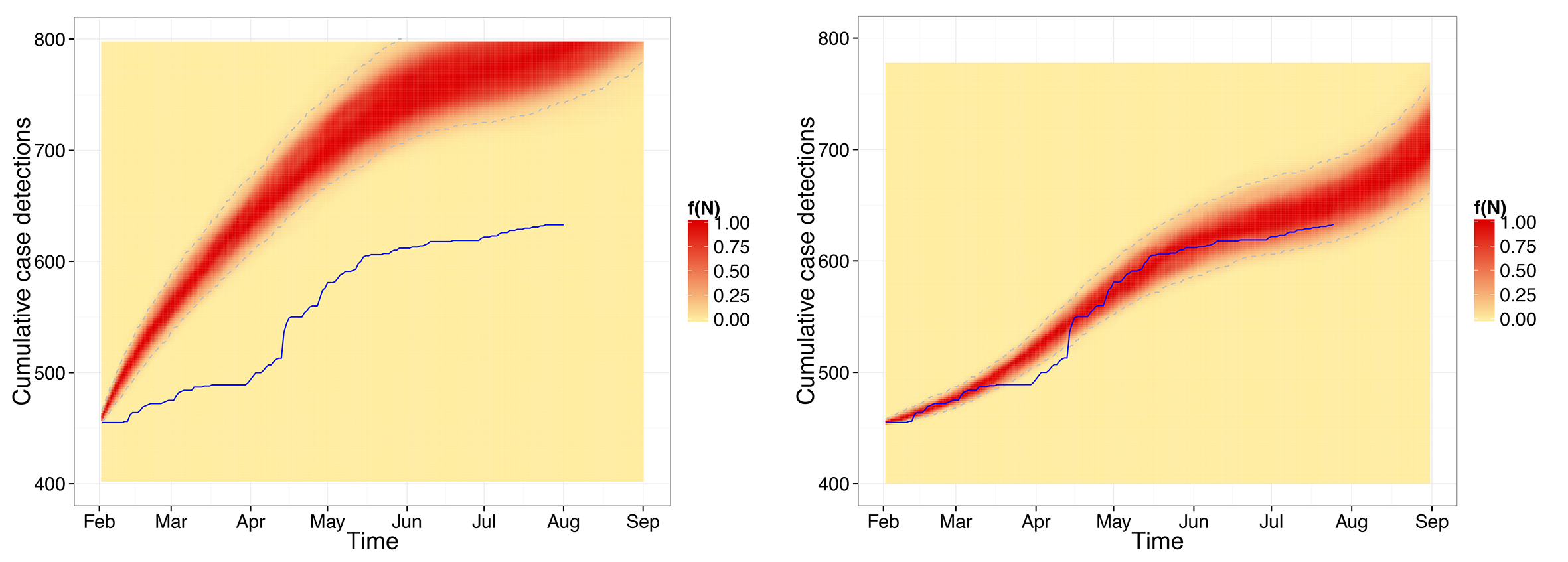}
\caption{\label{fig:ispred}6-month in-sample prediction obtained by simulating
forward from 1st February 2014, using parameter estimates generated as of
1st August 2014. Piecewise cubic spline (left) and square wave (right) seasonal
functions. Red smudge: the probability distribution over the predicted epidemic
trajectory with 95\% quantiles marked by the dotted lines. Blue line: Observed
cumulative cases. }
\end{figure}

\begin{figure}
\centering
\includegraphics[width=0.7\textwidth]{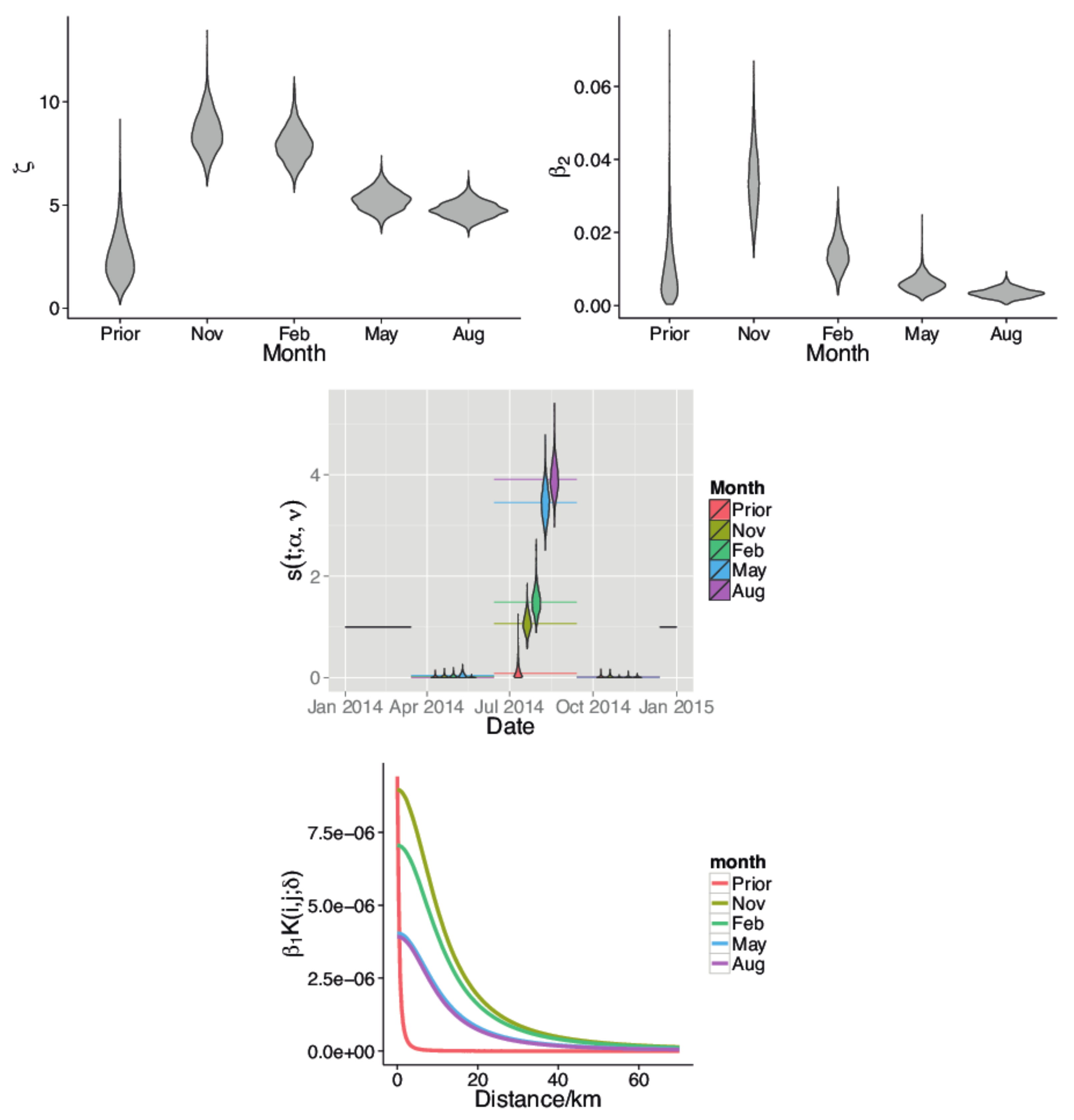}
\caption{\label{fig:betapost}Prior and posterior distributions at the four
analysis times for a) the probability of a NAIT movement transmitting an
infection $\beta_2$ (top left); b) the susceptibility of a dairy farm compared
to non-dairy $\zeta$ (top left); c) the median posterior seasonal step function
shown for each analysis time -- violin plots of the marginal posterior densities
for each analysis are shown for $\alpha_1$, $\alpha_2$, and $\alpha_3$,
corresponding to epochs 2, 3, and 4 respectively; d) the median environmental
transmission with distance $\beta_1 K(i,j;\delta)$ (bottom right). }
\end{figure}

Early in the analyses, it became apparent that the cubic spline seasonal
function was inadequate
to give sufficient posterior prediction power both for in-sample and
out-of-sample predictions. Figure \ref{fig:ispred}
presents the in-sample prediction for the August analysis, simulated from 1st
February. This clearly
shows that the cubic spline leads to a marked over-prediction in the size
of the epidemic, whereas the square
wave leads to a greatly superior in-sample prediction. However, the
discontinuities in the posterior distribution
introduced by the square wave presents significant difficulties in achieving
MCMC convergence for the period parameter $\nu$.
We therefore assumed $\nu=19/365$ from expert opinion \citep{Heath85}. The square
wave was used for
all subsequent results, and is represented in Figure \ref{fig:betapost}. The
model suggests that most
infection occurs in the third epoch, corresponding to mid-June to mid-September
for peak disease transmissibility.

The marginal prior and posterior distributions for the tick occurrence vector
$\bm{p}$ as of 1st August 2014 are summarised as median values in Figure \ref{fig:Cases-spatial}.
These posterior median values allow for population density due to the spatial
kernel, and therefore provide information on farm density adjusted regional transmission risk. 
On a national scale, the posterior medians show a similar tick distribution to the prior, as expected
from the distribution of epidemic cases and expert opinion. However, marked regional heterogeneity is
present in the North Island compared to the prior, reflecting a synthesis of
apparent tick prevalence from sampling and regional differences in disease
transmission.  We note that whilst the blood sampling data is the main influence on 
posterior tick occurrence, the effect of joint modelling with the epidemic data has a moderating effect
on individual TLA regions (Figure S1 in supporting information).

To assess the effect of time in our sequential analyses, Figure
\ref{fig:betapost} summarises the marginal posterior distributions for key
parameters in Equation \ref{eq:betaij}. A decrease in $\beta_2$ during 2014
indicates a decreased importance of the cattle movement network in transmitting
disease, concomitantly with a marked decrease in environmental transmission rate
after February 2014 as seen by the median of the posterior
$\beta_1K(i,j;\delta)$ spatial function. The posterior distributions for
$\zeta$, the susceptibility of dairy herds versus non-dairy shows two
populations of distributions. Here, dairy farms in the analyses prior to March
2014 are estimated to be approximately 8 times as susceptible as non-dairy
farms, whereas for the May and August 2014 this drops to approximately 5 times.
We notes that these graphs are both consistent with the flattening of the logged
cumulative case curve in Figure \ref{fig:Cases-spatial}. Interestingly, whereas
the model consistently estimates autumn and spring transmission to be
negibigble, an increasing trend is seen for the height of the spring seasonal
function. This indicates that although the overall apparent transmission rate is
tending to decrease with time, there is increased evidence for disease spread
being concentrated during the winter, given the acquisition of increasing
amounts of data. The infectious period (the time between a herd's infection and
detection events) is estimated consistently at 73 days (see supporting
information), consistent with the lag between seasonal transmission during the
winter, and the marked increase in case detection rate observed in the spring.

\begin{figure}
\centering
\includegraphics[width=0.7\textwidth]{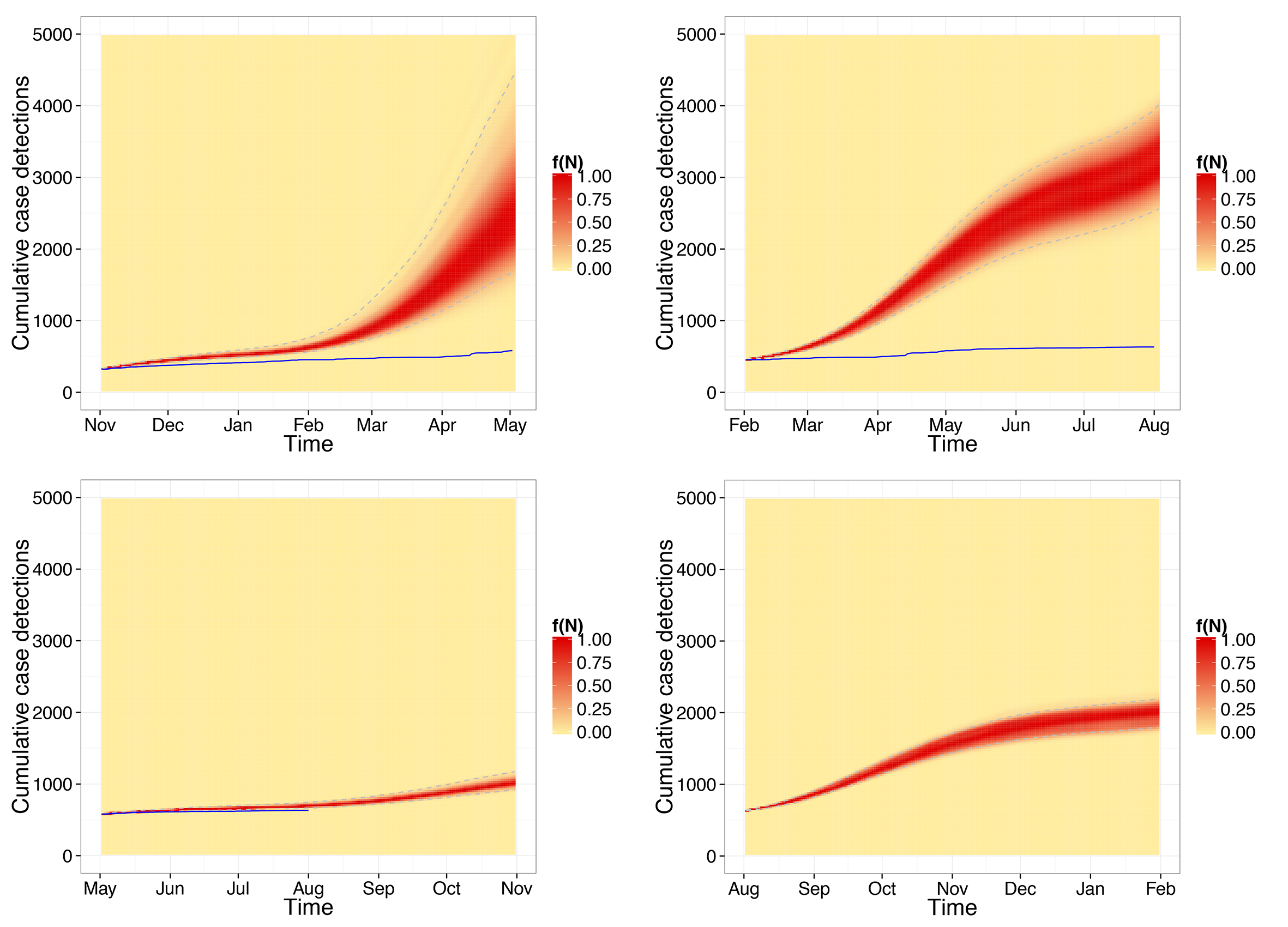}
\caption{\label{fig:oospred}6-month out-of-sample prediction of cumulative case
detections as of 1st November 2013 (top left),
1st February 2014 (top right), 1st May 2014 (bottom left) and 1st August 2014
(bottom right). Red smudge: the probability distribution over the predicted
epidemic trajectory with 95\% quantiles marked by the dotted lines. Blue curve:
The observed cumulative cases up to 1st August 2014.}
\end{figure}

A critical quantity in determining policy for a given disease outbreak is the
predicted size and extent of the epidemic. Figure \ref{fig:oospred} presents
6-month ahead predictions of cumulative numbers of cases detected based on the 3
analyses prior to August. Increases in the rate of case detections are predicted
for the autumn and spring periods.  This is due to the phase of the seasonal
function increasing the transmission rate during the summer and winter periods in
combination with the 73 day mean infection to detection time. In a sequential
setting, we evaluate out-of-sample predictive
ability by comparing the predictive distributions with the subsequent cumulative
case detections curve. An over-prediction is initially seen for both the
November and February analyses. For the May analysis this is much less apparent,
with the true number of case detections by August 1st lying on the 0.01
percentile of the predictive distribution. The improvement in this prediction
relates to the decreasing transmission rates and concentration of the infection
risk into the winter period as previously discussed.

\begin{figure}
\centering
\includegraphics[width=0.6\textwidth]{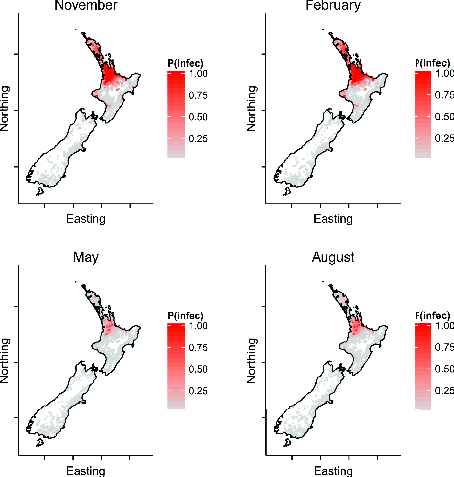}
\caption{\label{fig:new-case-distn}6-month predicted individual infection
probability, conditional on the observed epidemic as of 1st November 2013, 1st February 2014,
1st May 2014, and 1st August 2014. }
\end{figure}

The predicted spatial extent of the epidemic is represented by the probability
of individual herds becoming infected by 6 months ahead of the analysis times,
shown in Figure \ref{fig:new-case-distn}. These maps reflect the subsequent
spatial pattern of the epidemic as of 1st August (Figure \ref{fig:Cases-spatial}) 
and therefore are indicative of the likely extent of
the epidemic being confined to the north of the North Island in the medium term.
The large number of farms further South in the North Island account for the
significant number of subsequent cases occurring outside this area (Figure
\ref{fig:Cases-spatial}) event though individual infection probabilities are
low. We note, however, that predictions at the individual herd level are likely
to be inaccurate due to local model inadequacy and population data inaccuracies.

\section{Discussion}

The motivation behind this epidemic analysis was to provide rapid predictions
in response to a sudden incursion of the novel strain of vector-borne protozoan
\emph{T. orientalis} (Ikeda) in NZ cattle.
Little is known about the national level spatio-temporal dynamics of the
tick vector, and we have shown that a parsimonious approach which models the vector
population as a seasonal discrete-space latent risk surface is a viable option for 
serial prediction purposes. The results indicate an epidemic determined by both
vector presence and environmental transmission, with movements recorded in the NAIT
database having a low risk of propagating infection.

The main feature of our forecasting approach is to jointly model independent disease surveillance results
with epidemic data.  The \emph{a posteriori} estimates of the parameter vector $\bm{p}$ therefore
reflect a synthesis of static sample-based data and the dynamic epidemic data.  A more accurate interpretation
of $\bm{p}$ is therefore as a proxy for vector-driven disease transmission in 
each TLA region.  Of concern, however, is the 
level of spatial discretisation used for the BVD sampling data, though this was necessary to obtain anonymised data 
for the sampled farms.  We do not expect tick activity to be constant across TLA regions, nor do we expect 
a step change in tick activity across region borders.  Future research will therefore focus on integrating  
the recently characterised class of log Gaussian Cox processes for inference on continuous space risk surfaces 
given point data in preference to areal aggregations \citep{DigMorRowTay13}.

In our analyses, we have compared our models against subsequently observed data
using both in-sample and out-of-sample comparisons. Early in the analysis, in-sample
assessment of predictive performance quickly identified a strong preference for the square
wave seasonal function over the cubic spline, consistent with studies of the effect of humidity
and temperature on the activity of Ixodid ticks
\citep{ShEtAl89,KnRud82,St94,OgEtAl04,OsBr15}. We conclude therefore that in terms of disease transmission
these threshold effects are mimicked well by the square wave. In statistical terms the disadvantage of the square wave is its effect on the mixing 
quality of the MCMC, which is caused by the discontinuous nature of the posterior distribution with respect to changes in $\nu$.  A more elaborate 
cubic spline function, designed as a continuous approximation to the square wave, may well alleviate this particular difficulty albeit with the 
introduction of more parameters.  However, given that this dataset provides observations for only two replicates of an annual seasonal pattern, it is 
likely that a more complex model would exhibit a loss of statistical identifiability, again interfering with efficient model fitting.
We note also that non-identifiability between the phase ($\nu$) and infectious period ($b$) parameters is inherent to any periodic function, as the 
majority of infection
times are dictated by the season in which most disease transmission occurs. Thus
MCMC mixing issues are still apparent even for smooth functions.  Whilst further research is required to identify alternative seasonal functions, 
a promising approach to resolving this problem is to incorporate
climatic covariates into the analysis. Quantities such as vapour pressure, relative
humidity, and temperature at noon are all known to affect tick activity,
and estimating their effects as a hierarchical component within the
$h(j,t;\bm{\psi})$ function (Equation
\ref{eq:hFunction}) represents a straightforward extension to our model.

The out-of-sample predictive accuracy of our results changes markedly throughout
the epidemic as the spatiotemporal case detection data increase in volume. In
terms of the number of cases over time, the November and February analyses over
predict the number of future case detections by a large margin, with early
exponential growth dominating the rate of new case discovery far into 2014.
However, the subsequent observed cumulative case timeseries (Figure
\ref{fig:Cases-spatial}) shows a slowing of case detection rate in comparison to
exponential growth. This is only captured at the May analysis which predicts the
subsequent 3 month period far better with the acquisition of 126 new cases since
the February analysis. As such, our results are consistent with the tendency of
epidemic models to over predict numbers of cases, as is common due to unidentified
heterogeneity in the population \citep[for
example]{SrinivasaRao2006,Ong2010}. Features such as the timing of epidemic peaks,
seasonal effects, and spatial extent are however generally well identified.
Additionally, it is likely that a downward reporting bias is occurring as well
as a genuine slowing of the transmission rate as the cattle industry adapts to
the outbreak.

A striking difference between our results for NZ theileriosis and previous
results
from directly communicable diseases in animal populations is
the decay rate of the environmental transmission with distance.  
Previous studies in foot and mouth disease, and avian and equine influenza 
indicate that the majority of herd to herd transmission occurs within 5km
\citep{ChisSter2009, Cowled2009, JewEtAl2009a, MinhEtAl11}.
In contrast, our results suggest that environmental transmission of \emph{T.
orientalis} (Ikeda)
occurs over much greater distances.  We propose two possible explanations for this finding. 
Firstly, though ticks are relatively short ranging
arthropods (in comparison to
flying insect vectors), wildlife
hosts may be capable of translocating
infected ticks over long distances \citep{TenCh01}.  \emph{H. longicornis} has three lifecycle 
stages -- larva, nymph, adult -- with each stage feeding on a host \citep{Heath81}.  In principle, then, it may be possible 
for an adult, which has ingested \emph{T. orientalis} as a larva, to transmit the infection to a host in a remote location, after translocation at the 
nymphal stage.  We note that this may not require that the nymphal stage host be competent for \emph{T. orientalis}.
Secondly, a more plausible explanation
lies in the accuracy of NAIT-recorded movements with respect
to actual cattle movements around NZ, and also the accuracy of joining
Agribase\texttrademark~case identifiers to FOL and NAIT records. We note that
since the sparsity of the NAIT network is high, inaccuracies in georeferencing animal
movements will have a marked downward bias on $\beta_2$.  Additionally, NAIT
is a nascent movement recording system and we believe that, even though cattle
movement recording is mandatory, compliance may be low.  
In the absence of a national movement ban, it is therefore highly likely that
the apparent long-range spatial transmission
observed here is a result of reporting bias: the spatial transmission kernel
compensates for unrecorded animal movements,
with a corresponding downward bias on $\beta_2$.  That \emph{T. orientalis} (Ikeda) can be transmitted by 
the movement of tick-infested cattle is supported both by common sense and anecdotal evidence.  For example,
\citet{Islam2011} provide strong evidence for such a mode of infection through genetic typing of the pathogen.  However, even via this
mode of transmission, infection of a na\"{i}ve herd depends on local environmental conditions conducive to tick survival \citet{McFadden2011}.  We therefore conclude that risk of infection via incoming animal movements is determined by the geographical regions and time periods within which the vector population is active.

The availability of demographic and ecological data will always be the limiting
factor for detailed
dynamical models of disease. Whilst maintaining databases on livestock industry
demographics is commonly carried out
at the national level by government bodies, keeping pace with all possible
vector populations in the face of changing climate and habitats
is economically unfeasible. Bayesian data assimilation
and inference therefore provides a robust and rigorous solution for
quantitative decision support in disease response situations.

\section*{Acknowledgments}
The authors gratefully thank Andy McFadden, Mary van Andel, Daan Vink, and Kevin Lawrence at The Ministry for Primary Industries, and Robert 
Sanson at Asure Quality for supplying all data used in this study.  We thank Dr Allen Heath for his helpful discussions on arthropod vectors for 
\emph{Theileria}, and for supplying the prior information summarised in Figure \ref{fig:Cases-spatial}.

\clearpage

\appendix
\renewcommand\thesection{Appendix \Alph{section}}

%
%
%
%
%
%
%
%

\section{Preparation of NAIT movement network}

Demographic data for the New Zealand cattle herd population currently exist in a number of component databases. 
These require joining of records prior to analysis.  For our study, we have used Farms OnLine, NAIT, and Agribase.

Farms OnLine (FOL) is the government-owned agricultural property database.  Each record contains a unique identifier, owner contact information, geographic land parcel polygons (from which centroids are calculated), and presence/absence information for different animal types.   

The National Animal Identification and Tracking (NAIT) database records individual cattle and deer movements within New Zealand.  This is a mandatory system where each record represents an animal moving on a given date, between source and destination ``Persons In Charge of Animals'' (PICAs).  Furthermore, NAIT links each PICAs to FOL identifiers providing the opportunity to georeference animal movements for the purposes of joint network-spatial modelling.  

To begin, we represent the NAIT data as a dynamic network $A$ of cattle movements where nodes represent PICAs.  Let $N_{sr}(t)$ be a counting process describing the number of cattle moved between nodes $s$ and $r$ up to time $t$.  We estimate the mean directed edge frequency 
$$\hat{a}_{sr} = \frac{N_{sr}(t_1)-N_{sr}(t_0)}{t_1-t_0}$$
with $t_1-t_0$ the time interval represented by our NAIT extract.

To georeference this network, we map PICAs to FOL entities to obtain a new network $C$ with nodes represented by FOL entities.  However, a many-to-many relationship exists between PICAs and FOL entities.  Movements between PICAs linked to more than one FOL entity were therefore assumed to have occurred between a single pair of source/destination FOL locations with equal probability. The movement was therefore "distributed" between all possible pairs with equal weighting.  Furthermore, for consistency with the SID model we explicitly disallow loops. Thus the edge frequency for $C$ between FOL locations $i$ and $j$ is estimated as
\begin{equation*}
    \hat{c}_{ij} = \begin{cases}
							\sum_{s\sim i} \sum_{r\sim j} \frac{\hat{a}_{sr}}{m_s m_r} &  \mbox{if }i\ne j \\ 
							0 & \mbox{otherwise}
						\end{cases}
\end{equation*}
where $s\sim i$ denotes PICA $s$ being linked to FOL entity $i$ and $m_s = \sum_i 1[s \sim i]$ denotes the number of FOL entities associated with $s$.  Similarly for $r$ and $j$.

We note that no attempt to reflect uncertainty in the pair weighting is made in our study, though in principle this could be done either through a Dirichlet model, or via trans-dimensional MCMC as in \cite{Jewell2012}.

\subsection{Seasonal variation modelling}
For rapid likelihood-based inference and prediction, the seasonal function $s(t;\bm{\alpha},\nu)$
in Equation 3 (main paper) requires the following characteristics:  

\begin{enumerate}
\item \label{req:1}As part of the parameter estimation procedure, the definite integral of the
seasonal variation component $\int_{t_0}^t s(t; {\bm\alpha}, \nu)\,dt$ is
computed many thousands of times, for different parameter and $t$ values.
As such it is important that the integral is analytically tractable to
allow direct evaluation rather than costly quadrature computations. 
\item \label{req:2}The relative heights and depths of the spring/autumn and
summer/winter peaks and troughs, respectively, should be allowed to differ.
\item \label{req:3}It is desirable to set at least one of the seasonal peaks to a fixed value.  This aids statistical identifiability between $\bm{\alpha}$ and the baseline transmission parameters $\beta_1$ and $\beta_2$ (Equation 2, main paper) because the majority of the statistical information is concentrated into periods of high case incidence, corresponding to seasonal peaks.
 \end{enumerate}

Whilst at first a trigonometric function appears suitable for capturing seasonality, requirements \ref{req:2} and \ref{req:3} inevitably mean that \ref{req:1} is not satisfied.  To this end, we initially
modelled seasonal variation $s(t; {\bm\alpha}, \nu)$  using the following piecewise cubic formulation.
\begin{equation*}
  s(t; \bm{\alpha}, \nu) = \begin{cases}
    f_\text{spline}(t^\star; 0, 0.25, 1, \alpha_1) & 
    \text{if } 0    \le t^\star < 0.25 \\
    f_\text{spline}(t^\star; 0.25, 0.5, \alpha_1, \alpha_2) & 
    \text{if } 0.25 \le t^\star < 0.5  \\
    f_\text{spline}(t^\star; 0.5, 0.75, \alpha_2, \alpha_3) & 
    \text{if } 0.5  \le t^\star < 0.75 \\
    f_\text{spline}(t^\star; 0.75, 1, \alpha_3, 1) & 
    \text{if } 0.75  \le t^\star < 1 \\
  \end{cases}, 
\end{equation*}
where $t^\star = t + \nu - \lfloor t + \nu \rfloor$, with $t$ in years, and
where $f_\text{spline}(t; t_0, t_1, s_0, s_1)$ is the unique cubic passing
through $(t_0, s_0)$ and $(t_1, s_1)$ with $f'_\text{spline}(t_0) =
f'_\text{spline}(t_1) = 0$.  The parameter $0 \le \nu \le 0.5$ allows for phase adjustment of the seasonality. Note that to avoid identifiability
issues, the spring peak ($t^\star = 0$) is fixed at $1$ as the overall scale
is set by the $\bm\beta$ parameters (Equation 2, main paper).

\begin{figure}[H]
  \includegraphics[width=0.5\columnwidth]{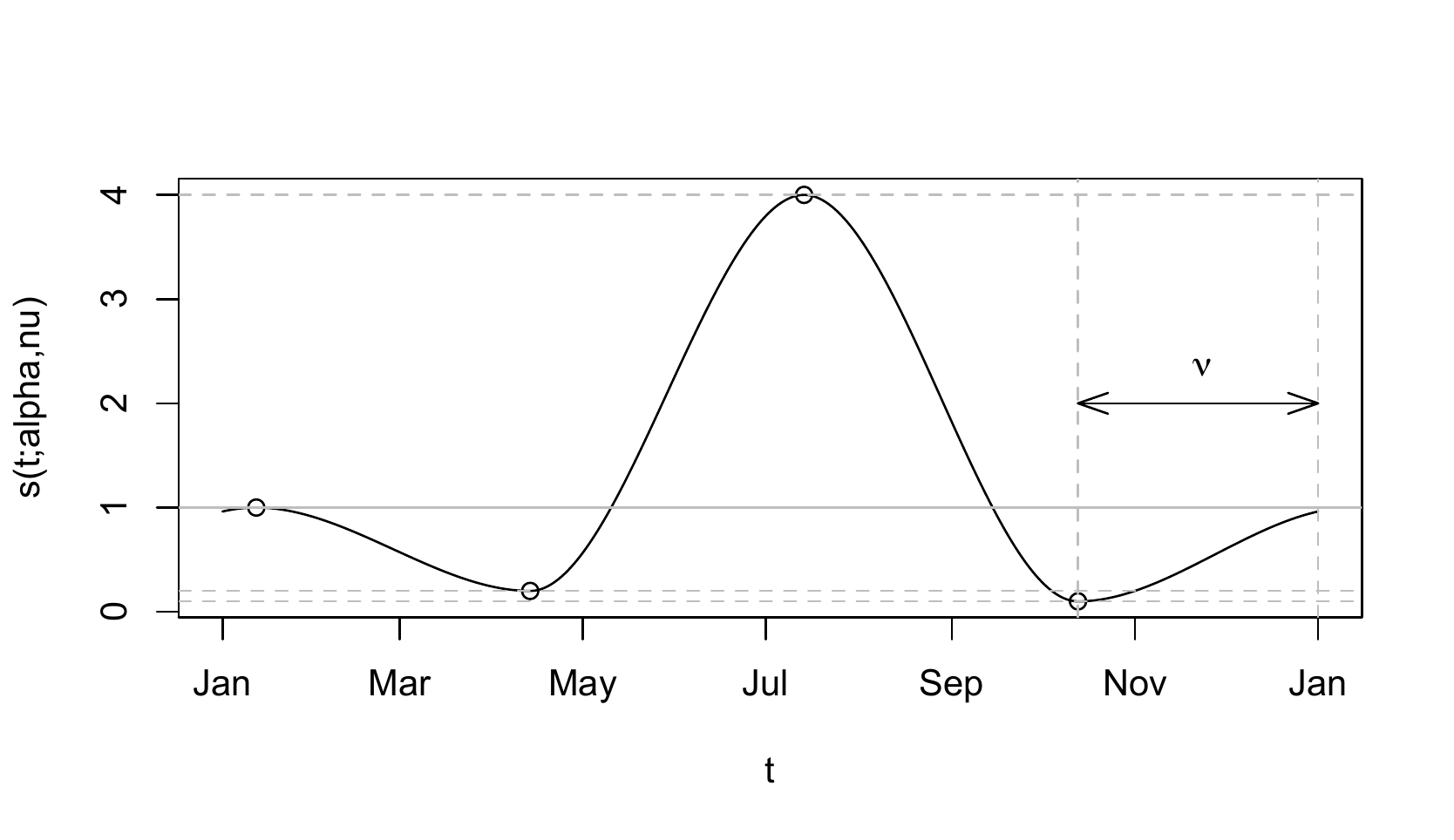}\hfill\includegraphics[width=0.5\columnwidth]{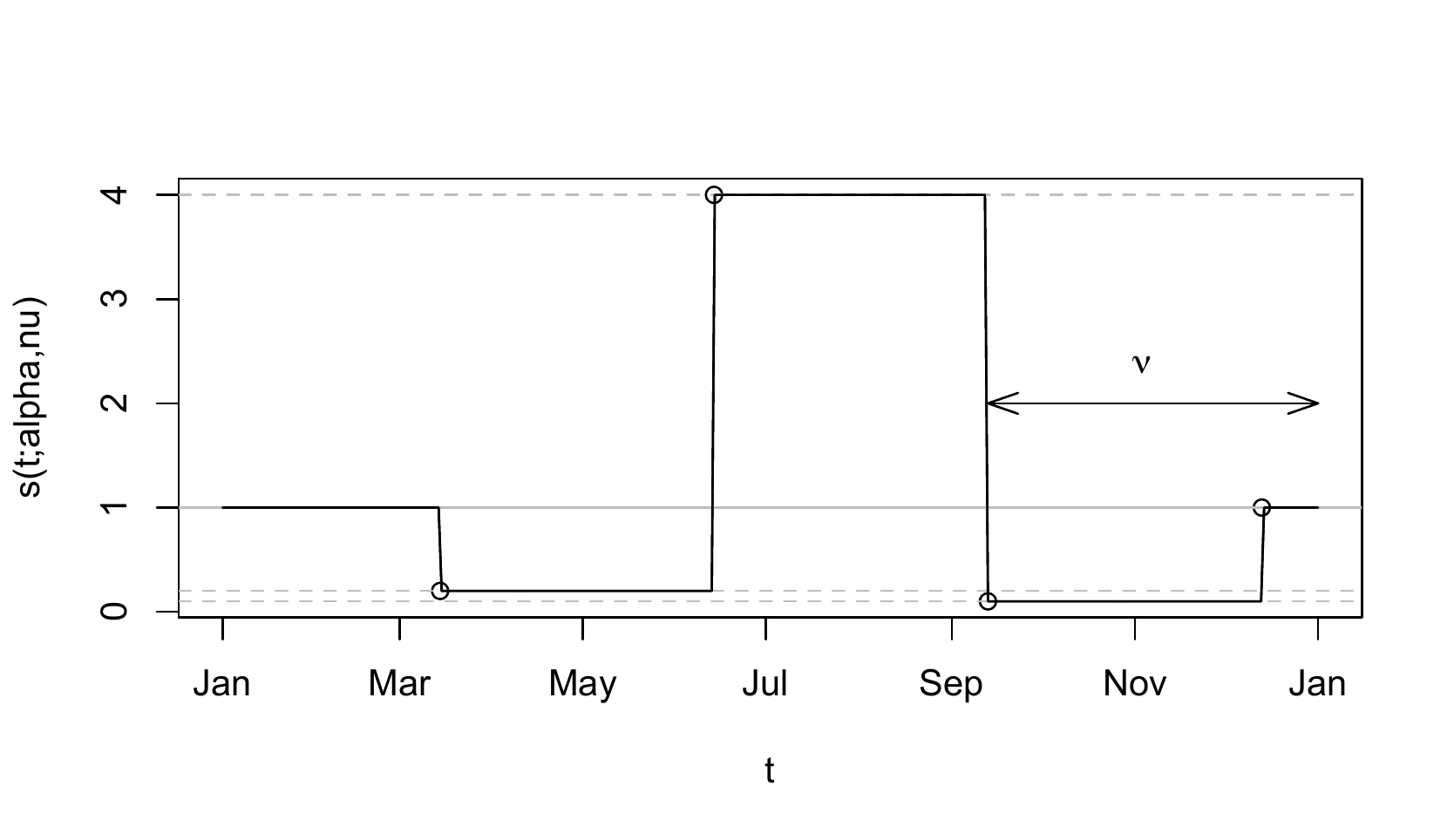}
  \caption{Sample spline (left) and square wave (right) seasonality functions.}
  \label{fig:seasonality}
\end{figure}

This function is, however, unable to specify neither prolonged periods of tick
activity or inactivity nor rapid threshold-like transitions between
regimes, and was subsequently rejected in favour of the square wave
function, (Equation 4, main paper)  which performed significantly better on the
data. See the discussion for more details. Figure~\ref{fig:seasonality}
shows example square-wave and cubic spline seasonality functions. 

\section{Statistical likelihood function}

We assume the epidemic and BVD sampling processes to be independent conditional on the underlying regional tick occurrence vector $\bm{p}$ as defined in the main text.  Our likelihood function falls into two parts: $L_E(\bm{\theta},\bm{I}|\bm{D})$ is the likelihood for the model parameters (including $\bm{p}$) and infection times conditional on the detection times, and $L_S(\bm{p}|\bm{X})$ is the likelihood for $\bm{p}$ given the sampling data $\bm{X}$.

To model the SID epidemic we use a continuous time inhomogeneous Poisson process setup.  The derivation of this likelihood function has been previously described (see for example \cite{AnBr00}).  

Let $T_{obs}$ be the analysis time of the epidemic, a time at which the epidemic may still be in progress.  At this time, we will have observed the times and identities of case detections represented by the $m_I$ dimensional vector $\bm{D}$.  The model parameters $\bm{\theta}=\{\bm{\alpha},\beta_1, \beta_2, \nu, \zeta, \bm{p}, b\}$ are unknown.  Furthermore, we do not observe the $M_I$ dimensional vector of infection times $\bm{I}$, which may include infection times corresponding to elements of $\bm{D}$ as well as those corresponding to undetected infections.  

Since $\bm{I}$ is not observed, we cannot write an explicit likelihood function for $\bm{\theta}$.  However, recalling the Gamma distributed infection to detection time (main text), we may write a conditional likelihood

\begin{eqnarray*}
L_E(\bm{\theta} | \bm{I},\bm{D}) & \propto & \prod_{j=1}{M_I} \left[\lambda_j(I_j^-)\right] \exp\left[\int_{I_\kappa}^{T_{obs}} \sum_{i\in\mathcal{P}} \lambda_j(t)dt\right] \\
& \times & \prod_{j=1}^{m_I} f_D(D_j-I_j) \\
& \times & \prod_{j=m_I+1}^{M_I} \left(1-F_D(D_j-I_j)\right)
\end{eqnarray*}
where $\mathcal{I}$ is the set of herds that have been infected up to the analysis time $T_{obs}$, $I_j$ and $D_j$ are the infection and detection times of the $j$th herd, and $\mathcal{P}$ is the set of individual herds comprising the entire population.  $\lambda_j(I_j^-)$ represents the infectious pressure on $j$ immediately before its infection, as described in the main text.  MCMC data augmentation methodology is then used to integrate over $\bm{I}$ in terms of both infection times and presence of undetected infections \cite{JewEtAl09c}.

To model the BVD sampling process, we use independent Binomial distributions for each TLA region $k=1,\dots,72$.  In many discrete-space statistical models, spatial dependency is explicitly included.  However, since the TLA regions are in general large, and the epidemic process explicitly accounts for spatial dependency, we choose not to incorporate this level of complexity.

Given sampling data $\bm{X}=\{\bm{n}, \bm{x}\}$ where $\bm{n}$ is the number of samples collected and $\bm{x}$ is the number of positive samples for \emph{Theileria orientalis} species, we have
$$L_S(\bm{p} | \bm{X}) \propto \prod_{k=1}^{72} p_k^{x_k}(1-p)^{n_k-x_k}$$

The joint (conditional) likelihood for the epidemic and BVD sampling process is therefore
$$L(\bm{\theta} | \bm{I}, \bm{D}) \propto L_E(\bm{\theta} | \bm{I}, \bm{D})L_S(\bm{p} | \bm{X})$$

In terms of inference, since $\bm{p}$ appears in both likelihood components (through $\lambda_j(t)$ for the epidemic), we allow its value to be determined not only by the BVD sampling, but also by the epidemic process.  $\bm{p}$ therefore represents a measure of tick-associated transmissibility in each TLA region.

\section{MCMC Convergence diagnostics}

\begin{figure}[H]
\centering
\includegraphics[width=.9\textwidth]{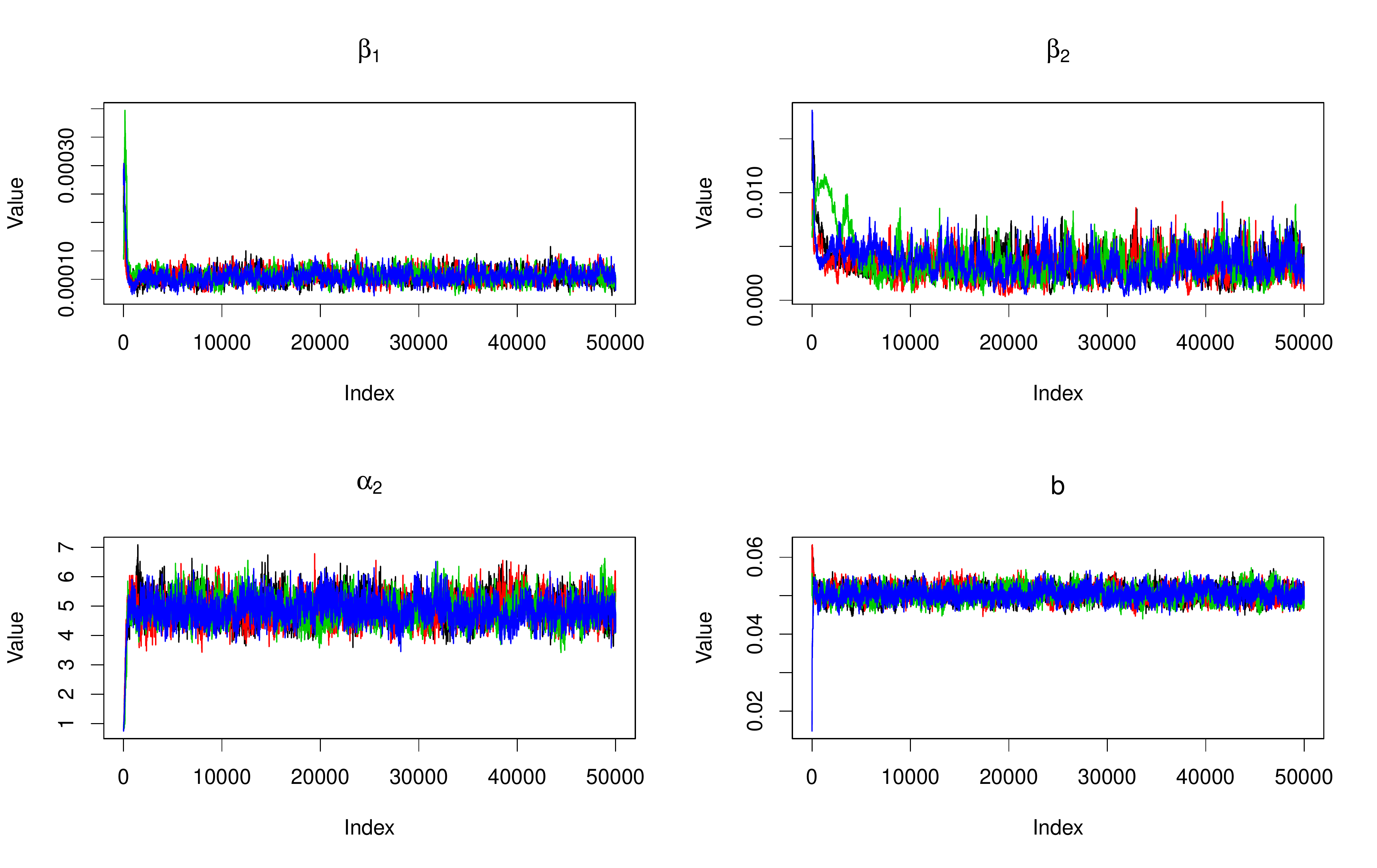}
\caption{\label{fig:augConvDiag.pdf}Superimposed traceplots for 4 parallel MCMC runs on the August dataset.  Graphs for $\beta_1$, $\beta_2$, $\alpha_2$, and $b$ are shown.}
\end{figure}

MCMC convergence was assessed by running 4 parallel chains starting at difference values for the parameters.  Superimposed timeseries plots were visually inspected.  The marginal traceplots for four parameters from the August analysis are shown in Figure \ref{fig:augConvDiag.pdf}.  These plots show the overall mixing quality of the algorithm to be satisfactory, though not perfect.  Importantly, the plots show that the assumption of the chains converging to the same limiting distribution is acceptable, confirmed by a Gelman-Rubin statistic of 1.001 \cite{BrGel98}.

\section{Priors}

Functional forms for the prior distributions used in our analysis were chosen to match the support of each parameter.  Typically, Gamma distribution are chosen for parameters describing rates (transmission rate, detection rate, etc), and Beta distributions are used for probability parameters (i.e. tick occurrence).  An exception to this is for parameters $\alpha_1$ and $\alpha_3$ which were given Beta distributions to ensure their definition as ``troughs'' in the seasonal function.  Prior distributions for $\bm{\alpha}$, $\beta_1$, $\beta_2$, $\zeta$, $\delta$, and $b$ are shown in Table \ref{tab:priorsepi}.  For $\bm{p}$, TLA regions are aggregated into ``high'', ``medium'', and ``low'' regions as shown in Figure 1 of the main text.  These are shown in Table \ref{tab:pprior}.

\begin{table}[H]
\caption{\label{tab:priorsepi}Prior distributions for parameters of the epidemiological model} 
\centering
\vspace{6pt}
\begin{tabular}{cc}
\textbf{Parameter} & \textbf{Distribution} \\
\hline
$\alpha_1$ & Beta(1,50) \\
$\alpha_2$ & Gamma(32, 8) \\
$\alpha_3$ & Beta(1,50) \\
$\beta_1$ & Gamma(4, 16000) \\
$\beta_2$ & Beta(2, 2) \\ 
$\zeta$ & Gamma(5, 2) \\
$\delta$ & Gamma(1, 1) \\
$b$ & Gamma(2.5, 50) \\
\hline
\end{tabular}
\end{table}

\begin{table}[H]
\caption{\label{tab:pprior}(Hyper)Parameters of the Beta($a,b$) distributions, together with the
median and 2.5\% and 97.5\% quantiles, used to encode information about spatial
vector risk for the three prior tick risk levels.}
\centering
\vspace{6pt}
\begin{tabular}{cccccc}
Prior vector risk & $a$ & $b$ & Median & 2.5\% quantile & 97.5\% quantile \\
\hline 
``high'' & 51 & 1 & 0.97 & 0.93 & 1.0 \\
``medium'' & 20 & 20 & 0.50 & 0.35 & 0.65 \\
``low'' & 1 & 50 & 0.014 & 0.00051 & 0.071 \\
\hline 
\end{tabular}
\end{table}

\section{Now-casting results}

\begin{figure}[H]
\centering
\includegraphics[width=0.8\columnwidth]{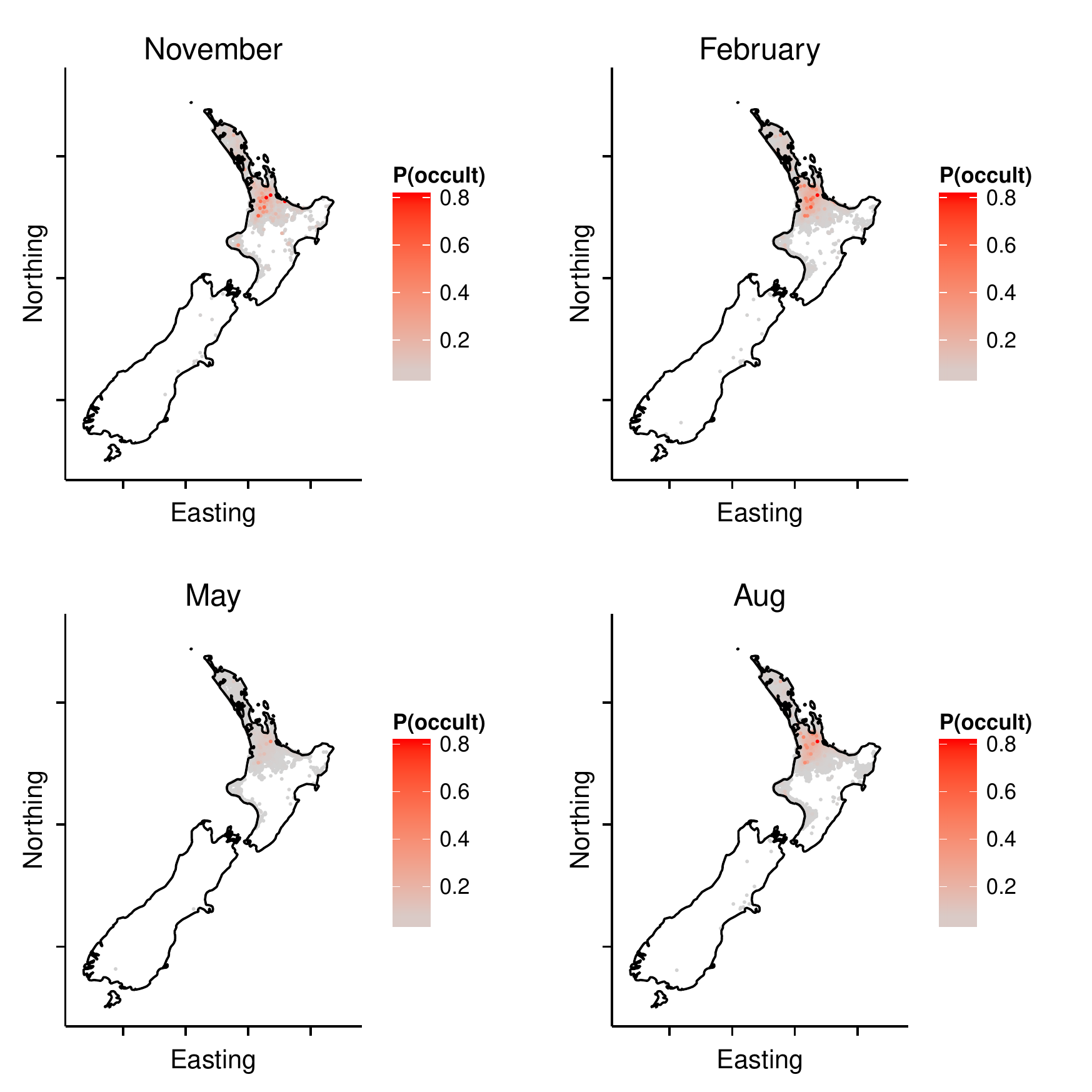}
\caption{\label{fig:occults}Spatial distribution of occult probabilities at each analysis timepoint.}
\end{figure}

This section contains further results on the current state of the epidemic, taken from the inference algorithm.  The maps in Figure \ref{fig:occults} provide a spatial representation of occult probability -- the probability that each presumed-susceptible farm is in fact an undetected infection.  Herds with a high probability of infection are confined to the high herd density region of northern Waikato and Auckland, reflecting the predominantly spatial nature of this epidemic.  These maps may be used to inform targeted surveillance by ranking farms in order of the most likely to be infected \cite{Jewell2012}.

\begin{figure}[H]
\centering
\includegraphics[width=0.5\columnwidth]{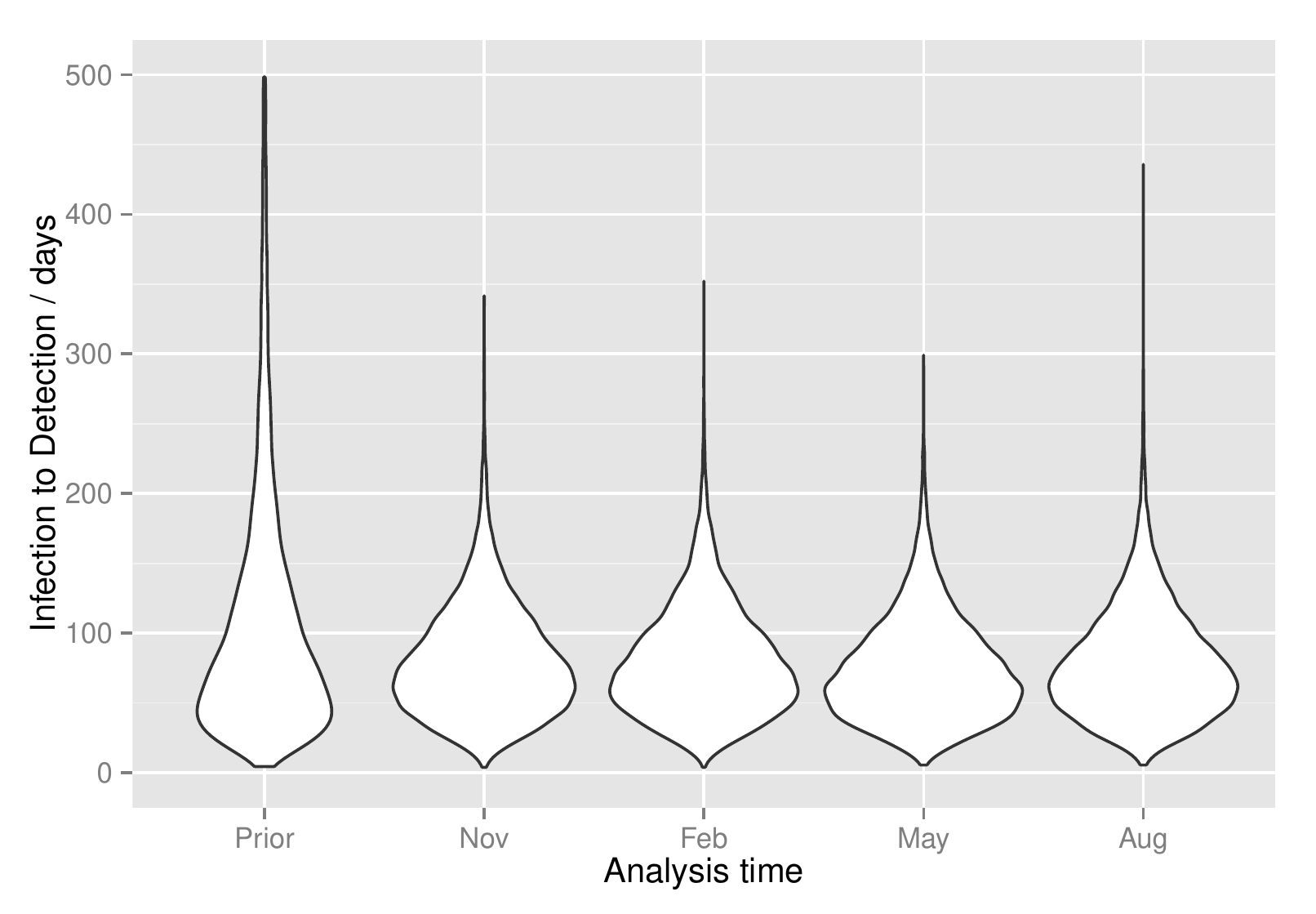}
\caption{\label{fig:bdist}Prior and posterior infectious period distributions for the 4 analysis times.}
\end{figure}

Figure \ref{fig:bdist} shows the posterior distributions for infection to detection time, a convolution of the infection to detection time model and the marginal posterior for $b$.  The results show a stability in the infectious period throughout the epidemic, with a reduction in posterior variance compared to the prior.  We note that for our data, the infectious period is essentially dictated by the phase of seasonal function.

\section{Effect of joint modelling on posterior tick occurrence}

\begin{figure}[H]
\centering
\includegraphics[width=0.35\columnwidth]{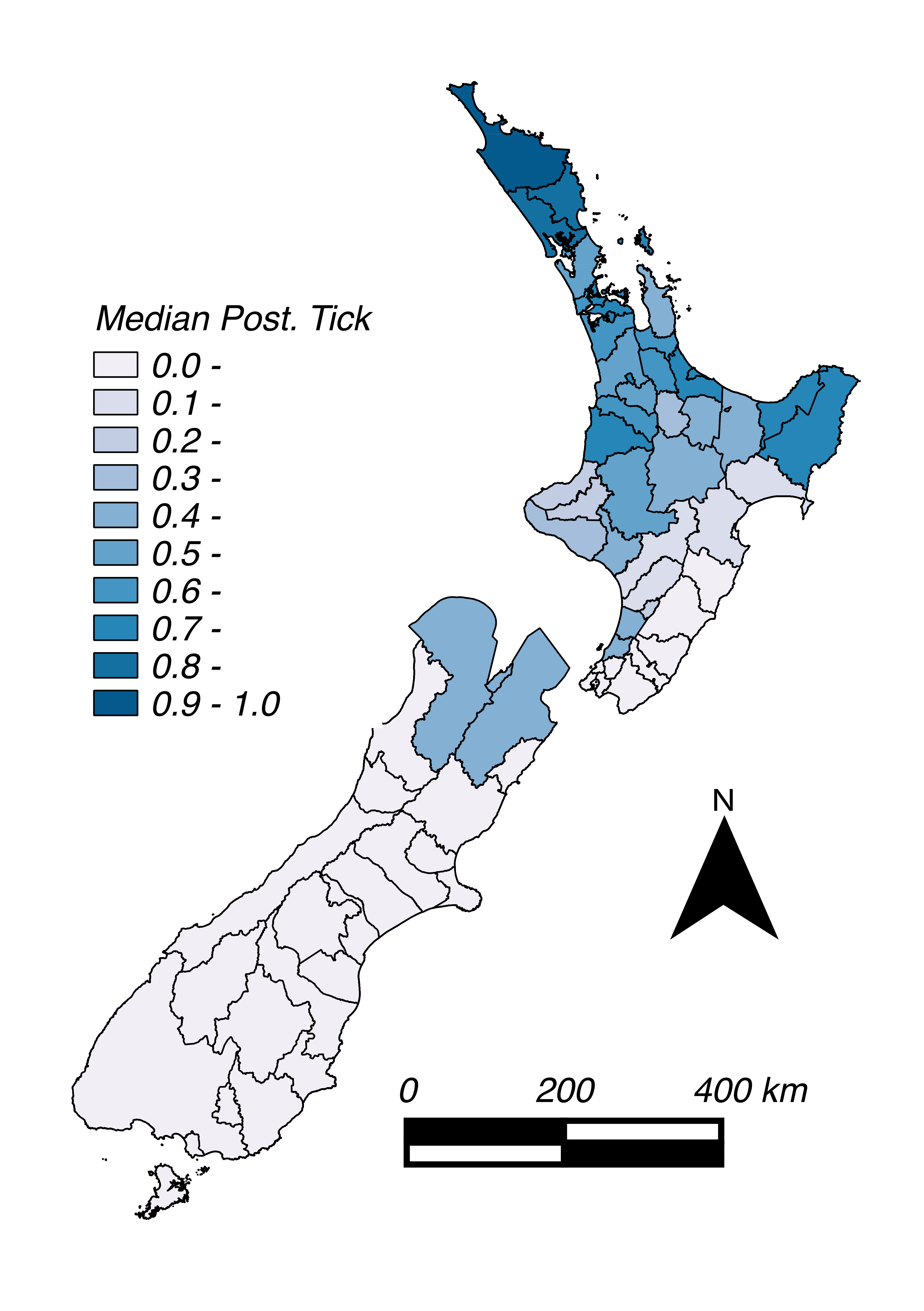} \vspace{1cm} \includegraphics[width=0.35\columnwidth]{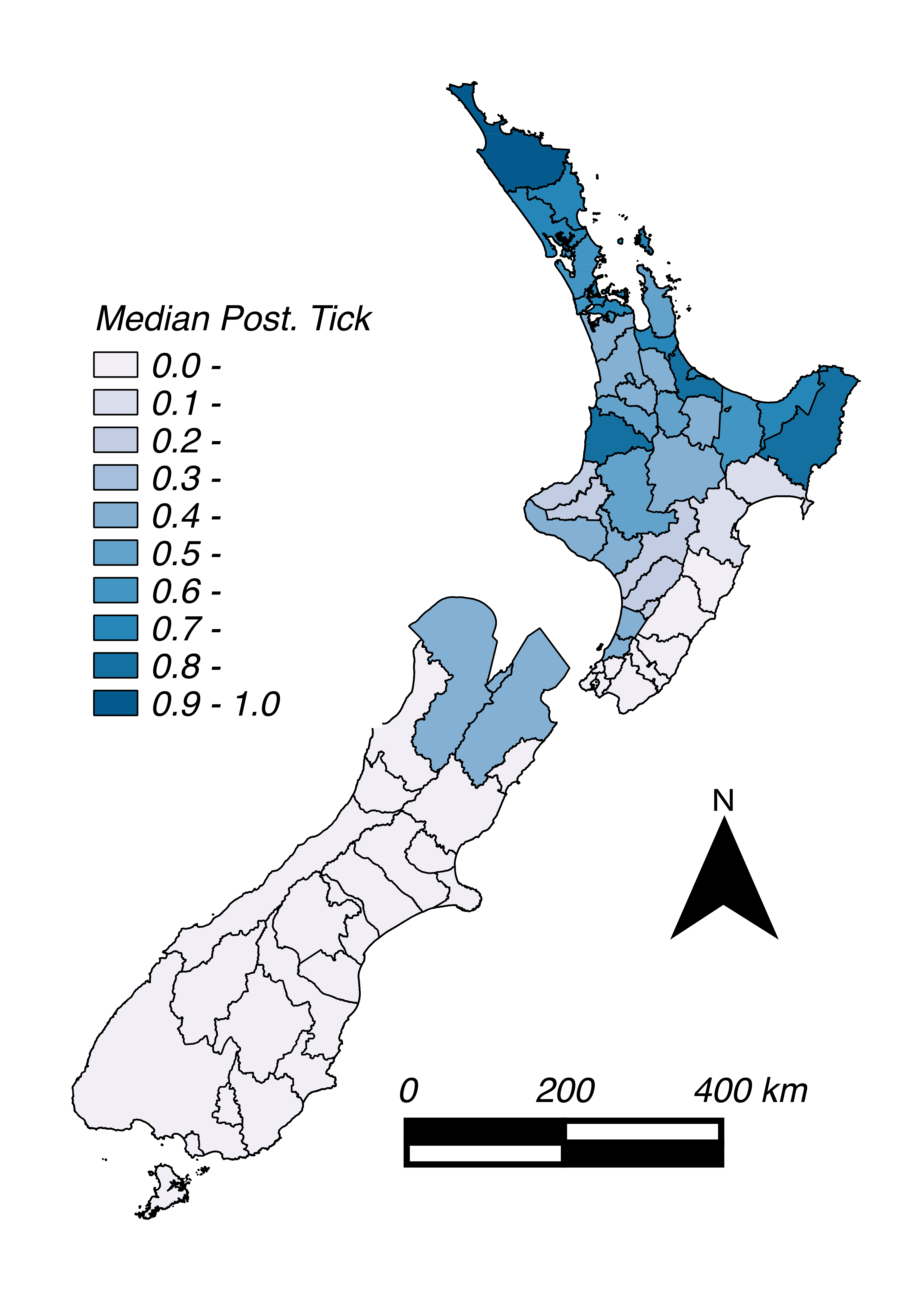}
\caption{\label{fig:samplingPostComp}A comparison of posterior TLA region tick occurrence for BVD sample data only (left) and joint  sample-epidemic data (right).}
\end{figure}

Figure \ref{fig:samplingPostComp} compares the posterior tick occurrence surface using just BVD sampling data, compared to that using the the joint sampling/epidemic model.  The overall pattern of tick activity is similar between the two models.  However, differences between the two maps in the high case density region of the North Island demonstrates how the innovation in joint modelling is required beyond marginal modelling of tick activity based on the sampling data alone.

\end{document}